\documentclass[reprint,showkeys,nofootinbib,amsmath,amssymb,aps,prl,]{revtex4-1}
\usepackage[marginal]{footmisc}
\usepackage{graphicx}% Include figure files
\usepackage{dcolumn}% Align table columns on decimal point
\usepackage{bm}% bold math
\usepackage{float}
\usepackage{wrapfig}
\usepackage{subfigure}
\usepackage{amsmath,amssymb}
\usepackage{booktabs}
\usepackage{url}
\usepackage{color}

\newcommand{\be}{\begin{equation}}
	\newcommand{\ee}{\end{equation}}
\newcommand{\ba}{\begin{eqnarray}}
	\newcommand{\ea}{\end{eqnarray}}

\begin{document}

	\title{Distinguishing gravity theories with networks of space-based gravitational-wave detectors}

	\author{Bo Mu$^{1,2}$}
	\email{mubo22@mails.ucas.ac.cn}
	\author{Zong-Kuan Guo$^{3,4,5}$}
	\email{guozk@itp.ac.cn}

    \affiliation{$^{1}$International Centre for Theoretical Physics Asia-Pacific, University of Chinese Academy of Sciences, 100190 Beijing, China}
	\affiliation{$^{2}$Taiji Laboratory for Gravitational Wave Universe, University of Chinese Academy of Sciences, 100049 Beijing, China}
	\affiliation{$^{3}$Institute of Theoretical Physics, Chinese Academy of Sciences, P.O. Box 2735, Beijing 100190, China}
	\affiliation{$^{4}$School of Physical Sciences, University of Chinese Academy of Sciences, No.19A Yuquan Road, Beijing 100049, China}
	\affiliation{$^{5}$School of Fundamental Physics and Mathematical Sciences, Hangzhou Institute for Advanced Study, University of Chinese Academy of Sciences, Hangzhou 310024, China}

	\begin{abstract}
		We propose a method for separating and detecting the non-tensor modes of stochastic gravitational-wave backgrounds (SGWBs) using networks of space-based gravitational-wave detectors. We consider four distinct data-reconstruction schemes for the co-inclination and anti-inclination orbital configurations of the LISA-Taiji network. We find that the co-inclination configuration offers its advantages over the anti-inclination one and can achieve signal-to-noise ratios up to 17.3 for the vector modes and 10.4 for the scalar modes with the energy density spectrum as $\Omega_{GW}^p(f)=10^{-12}$. Our method can be used to measure beyond-general-relativity polarization modes of SGWBs at mHz frequency band, opening a new avenue for testing alternative gravity theories.
	\end{abstract}
	
	\maketitle
	%%%%%%%%%%%%%%%%%%%%%%%%%%%%%%%%%%%%%%%%%%%%%%%%%%%%%%%%%%%%%%%%%
	\emph{Introduction}. 
	The stochastic gravitational-wave background (SGWB) is the superposition of a vast number of unresolved gravitational-wave signals, spanning both astrophysical and cosmological origins. On the astrophysical side, coalescing compact binaries (binary black holes and neutron stars), core-collapse supernovae~\cite{Ferrari1998}, rotating neutron stars (magnetars)~\cite{Rosado2012} and other stellar-remnant populations~\cite{Schneider2000} all can contribute the SGWB. Cosmological processes in the early Universe—quantum fluctuations amplified during inflation~\cite{Chen:2019xse,Madge:2020jjt,Inomata:2023zup,Liu:2023ymk,You:2023rmn,Wang:2023ost,Zhao:2023joc,Basilakos:2023xof,Basilakos:2023jvp,Jin:2023wri,Liu:2023pau,Liu:2023hpw,Firouzjahi:2023lzg,Franciolini:2023pbf,Ellis:2023oxs,Choudhury:2023fwk,Choudhury:2023hfm,tzerefos2024gravitationalwavesignaturesreheating,papanikolaou2024primordialblackholesinduced}, first-order phase transitions~\cite{Han:2023olf,Jiang:2023qbm,Fujikura:2023lkn}, and cosmic topological defects~\cite{Ellis:2023tsl,Wang:2023len,Lazarides:2023ksx,Kitajima:2023cek,Gouttenoire:2023ftk}—would also produce SGWB. Together, these components form a nearly isotropic, stationary signal that carries unique information about both the astrophysical population history and fundamental physics at energy scales far beyond those accessible on Earth. One can check these SGWB sources in this recent review~\cite{Christensen_2019}.

The search for SGWB has made substantial progress across different frequency ranges.  Ground-based interferometers (LIGO, Virgo, KAGRA) have now set stringent upper limits on the SGWB energy density in the $10–10^3$ Hz band ~\cite{Abbott2021}, while pulsar timing arrays (EPTA, NANOGrav, PPTA and CPTA) report evidence for a nanohertz-frequency common-spectrum stochastic signal~\cite{Agazie2023,Reardon2023,Antoniadis2023,Xu2023_CPTA_DR1},
whose physical origin remains under debate.  These results underscore the maturity of current GW observational techniques and the promise of probing cosmological and astrophysical sources of the SGWB.

Space-based detectors will extend this search into the millihertz window.  The Laser Interferometer Space Antenna (LISA)~\cite{ESA2017SciRD}, Taiji~\cite{WuHu2017Taiji,Ruan_2020_Taiji} and TianQin~\cite{li2024gravitationalwaveastronomytianqin} missions—each a three-spacecraft constellation with million-kilometer arms—are slated for launch in the 2030s.  Single-constellation analysis pipelines (e.g.\ \texttt{SGWBinner} ~\cite{Caprini2019}, \texttt{LISAtools} ~\cite{michael_katz_2024_10930980}) already demonstrate how time-delay-interferometry (TDI) channels can extract SGWB signals from instrumental noise.  Moreover, a joint network such as LISA–Taiji~\cite{Wang2021,Seto2020,Ruan_2020,Ruan:2019tje,Cai_2024} or LISA-TianQin~\cite{liang2025unveilingmulticomponentstochasticgravitationalwave}  yields multiple independent cross-correlation channels with distinct antenna patterns, provides better detectability for SGWB.

Beyond General Relativity (GR), generic four-dimensional metric theories admit up to six GW polarizations: the two transverse tensor modes of GR (“+” and “×”) ~\cite{Eardley1973,Will2014}, two transverse vector modes (“x” and “y”) ~\cite{Jacobson2004}, and two scalar modes (“breathing” and “longitudinal”) ~\cite{BransDicke1961,HassanRosen2012}. Vector–tensor models (e.g.\ Einstein–Æther) introduce the extra vector modes ~\cite{Jacobson2004}, scalar–tensor theories (e.g.\ Brans–Dicke) produce the breathing mode ~\cite{BransDicke1961,Chen_2025}, and massive-gravity or bimetric frameworks allow a longitudinal scalar polarization ~\cite{HassanRosen2012}. Each polarization has a unique overlap-reduction function and therefore a distinct signature in cross-correlated data streams ~\cite{Nishizawa2009}.

In ref.~\cite{OmiyaSeto2020}, Omiya \& Seto explore the prospects for isolating polarizations in the millihertz band using the LISA-Taiji network. By exploiting the special geometrical symmetry of the two triangular detectors, they construct a linear combination of cross-correlated data streams that algebraically cancels the standard tensor contribution, leaving only the vector and scalar signals.

In this paper, we propose a unified data-reconstruction method for SGWB polarization separation using space-based gravitational-wave detector networks. With LISA-Taiji network as the example, we calculate the six overlap-reduction functions(ORFs) for all six polarizations, reconstruct the cross-correlations to further separate vector/scalar polarization modes, and calculate the signal-to-noise ratios(SNRs) and sensitivity curves. Our results demonstrate the ability to isolate non-tensor components in the millihertz band, offering an opportunity to distinguish gravity theories. 
	
	%%%%%%%%%%%%%%%%%%%%%%%%%%%%%%%%%%%%%%%%%%%%%%%%%%%%%%%%%%%%%%%%%
	\emph{LISA and Taiji orbital configurations}.
    In this study, we investigate the performance of joint space-based gravitational wave observations using the Laser Interferometer Space Antenna (LISA)~\cite{ESA2017SciRD} and Taiji~\cite{WuHu2017Taiji} missions. Both missions are designed to probe the gravitational wave spectrum in the millihertz frequency band. The LISA constellation comprises three spacecraft forming an equilateral triangle with arm lengths of $2.5\times10^6$ km, trailing the Earth by approximately 20 degrees in its orbit, with the constellation plane inclined by 60 degrees relative to the ecliptic.

For the Taiji mission, we consider two different orbital configurations in this work. In both cases, the constellation leads the Earth by 20 degrees in orbit, but differs in the inclination of the constellation plane: one shares the same inclination as LISA (+60°), and the other adopts the opposite inclination (-60°). These are referred to as the co-inclination and anti-inclination configurations, respectively~\cite{Wang2021}. The constellations of these two Taiji configurations and LISA, along with their inclination are shown in Fig.~1.

\begin{figure}
    \centering
    \includegraphics[width=0.8\linewidth]{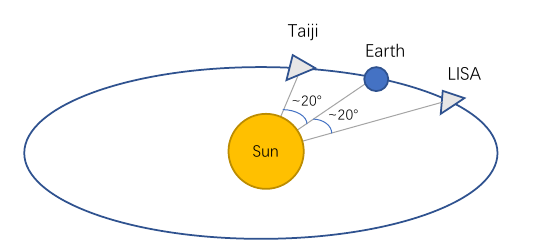}
    \includegraphics[width=0.8\linewidth]{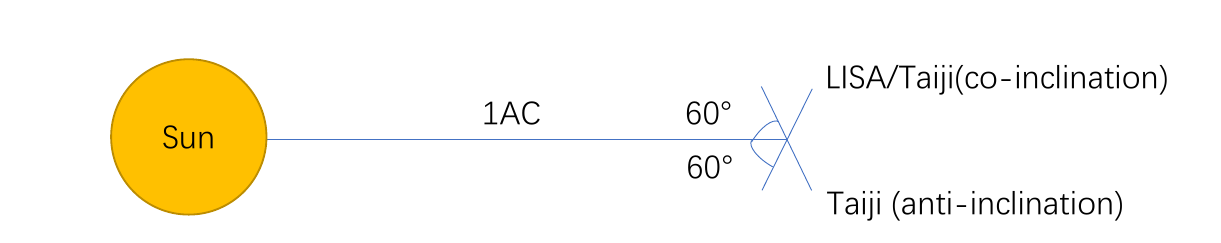}
    \caption{The upper picture is a schematic of the LISA and Taiji constellations relative to the Earth’s orbit.  The blue curve shows the Earth’s orbit around the Sun, with LISA trailing the Earth by $\sim20^\circ$ and Taiji leading by $\sim20^\circ$.The lower picture shows the inclination of the constellation planes relative to the ecliptic. }
    \label{fig:orbital}
\end{figure}

A third configuration, where Taiji is positioned at the same orbital location as LISA and shares its constellation plane, is not suitable for SGWB detection. In this colocated and coplanar setup, the two detectors may experience correlated instrumental or environmental noise due to their identical geometry and location. Such common noise undermines the fundamental assumption of statistical independence between detectors required for cross-correlation analyses. Therefore, this configuration is excluded from consideration in this study.

	\emph{Data channels and overlap reduction functions}.
    In order to detect the stochastic gravitational wave background (SGWB), it is essential to exploit the cross-correlation between independent data streams obtained from detectors. Each data channel in the frequency domain, denoted by \( s_a(f) \), consists of a gravitational wave signal \( h_a(f) \) and instrumental noise \( n_a(f) \),
\begin{equation}\label{eq:}
s_a(f) = h_a(f) + n_a(f).
\end{equation}
Here, the index \( a \in \{A, E, T\} \) labels the optimal TDI channels~\cite{TintoDhurandhar2014,TintoDhurandhar2005,deVine2010,WangNi2020,Liang2022,HartwigMuratore2022} of LISA, while \( b \in \{A', E', T'\} \) corresponds to the analogous channels of Taiji that appear later. The TDI channels \(A\), \(E\), and \(T\) are orthogonal linear combinations
of the Michelson channels \(X\), \(Y\), and \(Z\).
The Michelson channels \(X\), \(Y\), and \(Z\) are virtual equal-arm interferometers
formed by combining time-delayed one-way measurements between spacecraft in the
triangular constellation.
The \(A\), \(E\), and \(T\) channels are constructed to diagonalize the noise
covariance matrix, among which the \(A\) and \(E\) channels retain the dominant
gravitational-wave sensitivity, while the \(T\) channel is largely insensitive
to gravitational waves in the low-frequency regime.

The optimal TDI channels A, E, and T are defined as
\begin{equation}\label{eq:AET}
\mathrm{A} = \frac{Z - X}{\sqrt{2}}, \quad 
\mathrm{E} = \frac{X - 2Y + Z}{\sqrt{6}}, \quad 
\mathrm{T} = \frac{X + Y + Z}{\sqrt{3}}.
\end{equation}

We will introduce the TDI channels in the end of this section. 

In this article, we mark the tensor, vector and scalar polarization modes as \(\{T, V, S\} \). Assuming that instrumental noise between differant detectors are uncorrelated and that the SGWB is statistical, isotropic and Gaussian, the expectation value of the correlated signal pair simplifies to
\begin{equation}\label{eq:cross correlation}
\begin{aligned}
\langle s_a(f) \, s_b(f')^* \rangle &= \langle h_a(f) \, h_b(f')^* \rangle \\
&= \frac{1}{2}\delta(f - f') \sum_{p \in \{T, V, S\}} \gamma_{ab}^p(f)\, S_h^p(f),
\end{aligned}
\end{equation}
where \( p \) runs over the three possible gravitational wave polarization modes: tensor (T), vector (V), and scalar (S). The function \( \gamma_{ab}^p(f) \) denotes ORF, encoding the geometrical sensitivity of the detector pair to a stochastic background of polarization mode \( p \), and \( S_h^p(f) \) is the corresponding one-sided power spectral density. 
The strain power spectral density \( S_h^p(f) \) is related to the normalized energy density spectrum of the stochastic gravitational wave background. Under general relativity, for the tensorial modes, the dimensionless energy density \( \Omega_{\mathrm{GW}}^p(f) \) is given by
\begin{equation}\label{eq:energy density}
\Omega_{\mathrm{GW}}^p(f) = \frac{4\pi^2 f^3}{3(H_0/h)^2} S_h^p(f),
\end{equation}
where \( H_0/h \approx 3.24 \times 10^{-18}~\mathrm{s}^{-1} \) denotes the reduced Hubble parameter. It is important to note that this specific form of the relation between \( \Omega_{\mathrm{GW}}^p(f) \) and \( S_h^p(f) \) is derived within the framework of general relativity and applies strictly to tensor polarizations. In alternative theories of gravity, the energy density carried by vector or scalar modes could differ. For simplicity, we adopt this expression uniformly for all polarizations \( p \in \{T, V, S\} \) throughout this work.

In space-based interferometers, the dominant laser frequency noise is effectively canceled by TDI. After TDI suppression, the primary remaining noise sources are  the interferometric measurement system (IMS) noise and the test mass acceleration (acc) noise. The IMS noise characterizes
the displacement readout noise of the optical metrology system, while the
acceleration noise originates from residual non-gravitational forces acting on the
free-falling test masses. Their one-sided power spectral densities, adopted from Ref.~\cite{GotzeZaitsev2021}, are given by
\begin{equation}\label{noises}
\begin{aligned}
P_{\mathrm{IMS}}(f, P) &= P^2 \, \frac{\mathrm{pm}^2}{\mathrm{Hz}} \left[ 1 + \left( \frac{2\,\mathrm{mHz}}{f} \right)^4 \right] \left( \frac{2\pi f}{c} \right)^2, \\
P_{\mathrm{acc}}(f, A) &= A^2 \, \frac{\mathrm{fm}^2}{\mathrm{s}^4\,\mathrm{Hz}} \left[ 1 + \left( \frac{0.4\,\mathrm{mHz}}{f} \right)^2 \right] \left[ 1 + \left( \frac{f}{8\,\mathrm{mHz}} \right)^4 \right]
\\&\quad \times \left( \frac{1}{2\pi f} \right)^4 \left( \frac{2\pi f}{c} \right)^2,
\end{aligned}
\end{equation}
here $P$ and $A$ denote the dimensionless noise amplitude parameters for the IMS
and acceleration noise, respectively, following the standard LISA and Taiji noise
model conventions.
The corresponding A and E channel noise spectra are expressed as
\begin{equation}\label{overall noise}
\begin{aligned}
 N(f,A,P) &= N_{\mathrm{AA}}(f, A, P) = N_{\mathrm{EE}}(f, A, P) \\
&= 8 \sin^2\left( \frac{2\pi f L}{c}\right) 
\\    &\quad\times\Bigg\{ 
4 \left[ 1 + \cos\left( \frac{2\pi f L}{c} \right)+ \cos^2\left( \frac{2\pi f L}{c} \right) \right] 
\\    &\quad\times P_{\mathrm{acc}}(f, A) \\
&\quad + \left[ 2 + \cos\left( \frac{2\pi f L}{c} \right) \right] P_{\mathrm{IMS}}(f, P) 
\Bigg\}.
\end{aligned}
\end{equation}
We use fixed arm lengths \( L_{\mathrm{Taiji}} = 3.0 \times 10^9~\mathrm{m} \) and \( L_{\mathrm{LISA}} = 2.5 \times 10^9~\mathrm{m} \), which affect only the SNR calculation. The noise parameters are set to \( A = 3 \), \( P_{\mathrm{LISA}} = 10 \), and \( P_{\mathrm{Taiji}} = 8\), following Ref.~\cite{Robson2019,Wang2020}.

To compute the ORF between LISA and Taiji, we begin with the single-link response of a detector arm to a plane gravitational wave with polarization \(p'\)~\cite{WangNiHan2021}, where \(p' \in \{+, \times, x, y, b, L\}\). The frequency-domain response of the link from spacecraft \(i\) to \(j\) is given by

\begin{equation}\label{eq:response}
y^{p'}_{h,ij}(f) =
\frac{\hat{n}_{ij} \cdot \mathbf{e}_{p'} \cdot \hat{n}_{ij}}
{2\left(1 - \hat{n}_{ij} \cdot \hat{k}\right)}
\left[
e^{2\pi i f \left(L_{ij} + \hat{k} \cdot \mathbf{x}_i\right)}
-
e^{2\pi i f \left(\hat{k} \cdot \mathbf{x}_j\right)}
\right],
\end{equation}
where \( \hat{n}_{ij} \) is the unit vector pointing from spacecraft \(i\) to \(j\), \( L_{ij} \) is the arm length, $\mathbf{x}_i$ and $\mathbf{x}_j$ denote the barycentric position vectors
of spacecraft $i$ and $j$, and \( \mathbf{e}_{p'} \) is the polarization tensor for mode \(p'\). The gravitational wave propagation direction is given by
\begin{equation}\label{eq:propagation direction}
\hat{k} = -(\cos \lambda \cos \theta, \, \sin \lambda \cos \theta, \, \sin \theta),
\end{equation}
where \( \lambda \) and \( \theta \) are the ecliptic longitude and latitude of the source in the heliocentric barycentric coordinate system. And the polarization tensor $e_{p'}$ is expressed as

\begin{equation}\label{polarization modes}
\begin{aligned}
e_{+} &\equiv O_{1}
\begin{pmatrix}
1 & 0 & 0\\
0 & -1 & 0\\
0 & 0 & 0
\end{pmatrix}
O_{1}^{T},
&\quad
e_{\times} &\equiv O_{1}
\begin{pmatrix}
0 & 1 & 0\\
1 & 0 & 0\\
0 & 0 & 0
\end{pmatrix}
O_{1}^{T}\times i,\\[1ex]
e_{b} &\equiv O_{1}
\begin{pmatrix}
1 & 0 & 0\\
0 & 1 & 0\\
0 & 0 & 0
\end{pmatrix}
O_{1}^{T},
&\quad
e_{L} &\equiv O_{1}
\begin{pmatrix}
0 & 0 & 0\\
0 & 0 & 0\\
0 & 0 & 1
\end{pmatrix}
O_{1}^{T},\\[1ex]
e_{x} &\equiv O_{1}
\begin{pmatrix}
0 & 0 & 1\\
0 & 0 & 0\\
1 & 0 & 0
\end{pmatrix}
O_{1}^{T}\times\frac{1}{2},
&\quad
e_{y} &\equiv O_{1}
\begin{pmatrix}
0 & 0 & 0\\
0 & 0 & 1\\
0 & 1 & 0
\end{pmatrix}
O_{1}^{T}\times i.
\end{aligned}
\end{equation}

with
\begin{equation}\label{O1}
\quad
O_{1} = 
\begin{pmatrix}
\sin\lambda 
  & \;-\;\cos\lambda\sin\theta
  & -\,\cos\lambda\cos\theta \\[1ex]
-\,\cos\lambda 
  & -\;\sin\lambda\sin\theta
  & -\,\sin\lambda\cos\theta \\[1ex]
  0
  & \cos\theta
  & -\,\sin\theta
\end{pmatrix}.
\end{equation}
In the case of an isotropic SGWB, all polarization states are assumed to be equally probable and uniformly distributed over the sky. Consequently, the polarization angle becomes an unobservable degree of freedom in our measurement. For simplicity—and without loss of generality—we therefore set the polarization angle to zero. As a result, no explicit dependence on the polarization angle appears in the expressions for the polarization tensors or overlap reduction functions used throughout this work.

\begin{figure}
\centering
    \includegraphics[width=0.8\linewidth]{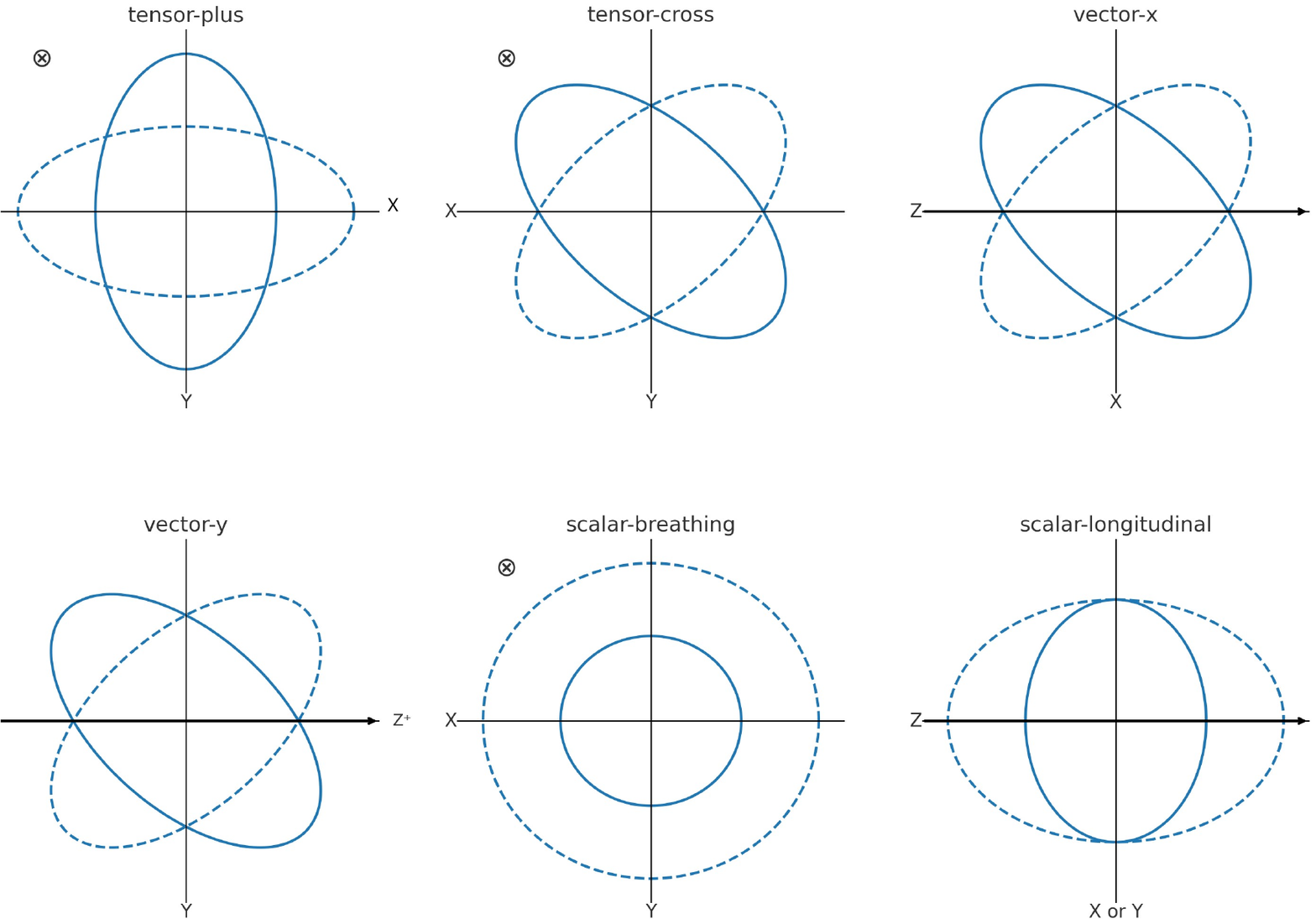}
    \caption{Top‐down deformation of a unit ring by the six independent gravitational‐wave polarizations: tensor-plus and tensor-cross in the X–Y plane; vector-x in X–Z and vector-y in Y–Z; scalar “breathing” in X–Y; and scalar “longitudinal” along Z (identical when viewed from X or Y). Solid and dashed curves show the two quadrature phases.}
    \label{fig:polarization}
\end{figure}

Figure \ref{fig:polarization} provides a unified “top-down” visualization of the six independent polarization modes that a gravitational wave may exhibit.  In each panel we begin with a unit ring of test particles (solid and dashed curves correspond to two quadrature phases) and show how that ring is deformed by the wave in the plane indicated by the central cross.  The upper left and upper middle panels display the familiar tensor‐plus and tensor‐cross modes in the X–Y plane—respectively stretching and squeezing along the X and Y axes or along the diagonals—while the upper right and the lower left panels illustrate the two vector modes, which produce figure‐eight distortions in the X–Z and Y–Z planes.  The lower middle panel shows the scalar “breathing” mode, in which the entire ring expands and contracts radially in the X–Y plane, and the lower right panel shows the scalar “longitudinal” mode, in which particles oscillate along the wave’s direction of travel (the Z axis), appearing identical whether viewed from the X or Y axis.  This collection of deformations makes explicit the distinct signatures one would seek in a detector network when testing General Relativity’s tensor‐only prediction against alternative theories that admit additional vector or scalar polarizations.

%The projection factor in the denominator accounts for the light travel delay due to wavefront orientation, while the exponential terms encode the phase shift at the spacecraft endpoints.

To construct the optimal interferometric observables, we start with the first-generation TDI Michelson-X combination. This configuration synthesizes a virtual equal-arm interferometer using delayed one-way phase measurements along specific spacecraft paths. The response of the Michelson-X channel to a gravitational wave with polarization \(p\) in the frequency domain is expressed as a linear combination of single-link responses:
\begin{equation}\label{TDI X}
\begin{aligned}
F_{X}^{p'}(f) &= (-\Delta_{21} + \Delta_{21} \Delta_{13} \Delta_{31}) y^p_{h,12} 
+ (-1 + \Delta_{13} \Delta_{31}) y^{p'}_{h,21} \\
&\quad + (\Delta_{31} - \Delta_{31} \Delta_{12} \Delta_{21}) y^{p'}_{h,13} 
+ (1 - \Delta_{12} \Delta_{21}) y^{p'}_{h,31},
\end{aligned}
\end{equation}
where \( \Delta_{ij} = e^{-2\pi i f L_{ij}} \) is the frequency-domain time-delay operator for the light travel time along the arm from spacecraft \(i\) to \(j\), and \( y^{p'}_{h,ij} \) is the response of the individual link as defined earlier. Similar expressions apply for the Y and Z channels by cyclic permutation of spacecraft indices. With Eq.~\eqref{eq:AET}, the response of the optimal TDI channels can be obtained.

The ORF quantifies the sensitivity of a pair of detectors to a stochastic gravitational wave background with polarization \(p'\). It accounts for the relative orientation, spatial separation, and frequency-dependent response of the detectors to a gravitational wave arriving from all directions. For TDI channels \(a\) and \(b\) from different detectors, the ORF is given by
\begin{equation}\label{eq:ORF}
\gamma_{ab}^{p'}(f) = \frac{1}{4\pi} \int \mathrm{d}\mathbf{n} \,
F^{p'}_{a}(f, \mathbf{n}) \, F^{p'}_{b}(f, \mathbf{n})^*,
\end{equation}
where \(F^{p'}_a(f, \mathbf{n})\) and \(F^{p'}_b(f, \mathbf{n})\) denote the detector responses to polarization \(p\) for a plane wave incident from direction \(\mathbf{n}\), and the integral is taken over the full sky. The ORF encodes the coherence between data streams from LISA and Taiji for a given polarization and frequency, and plays a central role in determining the cross-correlation signal strength. In practice, \(\gamma_{ab}^{p'}(f)\) must be evaluated numerically using the full detector response formalism introduced earlier. With \(\gamma_{ab}^{p'}(f)\), we can calculate  \(\gamma_{ab}^{p}(f)\) as follow,
\begin{equation}\label{eq:ORF sum}
\begin{aligned}
\gamma_{ab}^{T}(f) &= \gamma_{ab}^{+}(f) + \gamma_{ab}^{\times}(f), \\ 
\gamma_{ab}^{V}(f) &= \gamma_{ab}^{x}(f) + \gamma_{ab}^{y}(f), \\
\gamma_{ab}^{S}(f) &= \gamma_{ab}^{b}(f) + \gamma_{ab}^{L}(f),
\end{aligned}
\end{equation}
here $\gamma_{ab}^{T}$, $\gamma_{ab}^{V}$, and $\gamma_{ab}^{S}$ denote the effective
overlap reduction functions for tensor, vector, and scalar sectors, respectively.
These definitions correspond to an unpolarized and statistically isotropic SGWB,
for which different polarization states within the same sector are assumed to be
independent, equally populated, and uncorrelated. In this case, the total response
is given by the sum of the individual polarization contributions at the level of
the power spectral density
\cite{AllenRomano1999,Nishizawa2009,Christensen:2018iqi}.

\emph{Data reconstruction}.
To extract specific polarization components of the stochastic gravitational wave background, we construct linear combinations of cross-correlated data from multiple TDI channels. We use the \(A\) and \(E\) channel of LISA and  \(A'\) and \(E'\) channel Taiji to do the cross-correlation, thus we have four cross-correlation channels: \((A, A'), (A, E'), (E, A'),(E, E')\) to use. To isolate one polarization mode, we need to cancel the other two modes. The reconstruction procedure needs three data channels, so four different combinations for reconstruction can be accessible. Each combination is designed to isolate a particular polarization by exploiting differences in the overlap reduction functions between channels. 

We separate the scalar polarization component with \((A, A'), (A, E'), (E, A')\) channels in this section as an example. Two Taiji configurations, a co-inclination one and an anti-inclination one, use the same procedure. We define the following combination of cross-correlated signals:
\begin{equation}\label{mu S}
\begin{aligned}
\mu_S &= (\gamma_{AE'}^V \gamma_{EA'}^T - \gamma_{EA'}^V \gamma_{AE'}^T) \, s_A s_{A'}^* \\
&\quad + (\gamma_{EA'}^V \gamma_{AA'}^T - \gamma_{AA'}^V \gamma_{EA'}^T) \, s_A s_{E'}^* \\
&\quad + (\gamma_{AA'}^V \gamma_{AE'}^T - \gamma_{AE'}^V \gamma_{AA'}^T) \, s_E s_{A'}^*,
\end{aligned}
\end{equation}
 where $\gamma$ is the ORF, treated as constant coefficient, and s is the strain data from TDI channel. Under the assumption that the background is dominated by instrumental noise (\(|h_a| \ll |n_a|\)), the expectation value of \(\mu_S\) isolates the scalar signal as
\begin{equation}\label{mu S expectation value}
\begin{aligned}
\langle \mu_S \rangle &=\frac{1}{2}[ (\gamma_{AE'}^V \gamma_{EA'}^T - \gamma_{EA'}^V \gamma_{AE'}^T) \gamma_{AA'}^S S_h^S(f) \\
&\quad + (\gamma_{EA'}^V \gamma_{AA'}^T - \gamma_{AA'}^V \gamma_{EA'}^T) \gamma_{AE'}^S S_h^S(f) \\
&\quad + (\gamma_{AA'}^V \gamma_{AE'}^T - \gamma_{AE'}^V \gamma_{AA'}^T) \gamma_{EA'}^S S_h^S(f) ].
\end{aligned}
\end{equation}
In this expectation value, the contributions from $S_h^T(f)$ and $S_h^v(f)$ vanish by virtue of the data reconstruction, leaving only the scalar–polarization term $S_h^S(f)$. Consequently, any SGWB signal appearing in the reconstructed data $\mu_S$ must be due to the scalar mode. The reconstructed vector–mode data $\mu_V$ behaves analogously. This reconstruction procedure thus allows us to isolate and detect a single SGWB polarization mode directly, offering a clear way to distinguish gravity theories by comparing their predicted polarization content with observations.

The variance of the reconstructed scalar combination \(\mu_S\) is determined by the noise cross-correlations. With $
\langle n_a(f) \, n_a(f')^* \rangle
= \frac{1}{2}\delta(f - f') N_{aa}(f)
$, neglecting signal contributions, the variance is given by
\begin{equation}\label{variance}
\begin{aligned}
\sigma_{\mu_S}^2(f) &= \frac{1}{4} \big[ 
(\gamma_{AE'}^V \gamma_{EA'}^T - \gamma_{EA'}^V \gamma_{AE'}^T)^2 \\
&\quad + (\gamma_{EA'}^V \gamma_{AA'}^T - \gamma_{AA'}^V \gamma_{EA'}^T)^2 \\
&\quad + (\gamma_{AA'}^V \gamma_{AE'}^T - \gamma_{AE'}^V \gamma_{AA'}^T)^2 
\big] N(f) N'(f),
\end{aligned}
\end{equation}
where \(N(f)\) and \(N'(f)\) denote the noise power spectral densities for the LISA and Taiji channels, respectively. With this, the total SNR for the scalar mode detection is obtained by integrating over frequency:
\begin{equation}\label{eq:SNR}
\mathrm{SNR}^2 = \int_{f_{\mathrm{low}}}^{f_{\mathrm{high}}} \frac{\langle \mu_S(f) \rangle^2}{\sigma_{\mu_S}^2(f)} \, df.
\end{equation}
The total SNR for the scalar component of the stochastic gravitational wave background can be written as
\begin{equation}\label{eq:true SNR}
\mathrm{SNR}^2 = \left( \frac{3(H_0/h)^2}{4\pi^2} \right)^2 T_{\mathrm{obs}}  \left[ \int_{f_{\mathrm{low}}}^{f_{\mathrm{high}}} df \, \frac{ \left[ \Gamma^S(f)\, \Omega_{\mathrm{GW}}^S(f) \right]^2 }{ f^6 N(f) N'(f) } \right],
\end{equation}
where \( T_{\mathrm{obs}} \) is the total observation time, \( \Omega_{\mathrm{GW}}^S(f) \) is the normalized energy density spectrum of the scalar-mode gravitational wave background, and \( N(f) \), \( N'(f) \) denote the noise spectral densities of the LISA and Taiji channels, respectively. We take $f_{\mathrm{low}}= 10^{-4}\mathrm{Hz}$ and $f_{\mathrm{high}}=10^{-1}\mathrm{Hz}$ as our frequency range, since this range includes the sensitive range of both LISA and Taiji.

The effective scalar-mode overlap reduction function \( \Gamma^S(f) \) captures the geometric sensitivity of the multi-channel cross-correlation configuration to scalar polarization, and is defined by
\begin{equation}\label{eq:Reconstructed ORF}
\Gamma^S(f) = \frac{
\alpha_1(f)\, \gamma_{AA'}^S(f)
+ \alpha_2(f)\, \gamma_{AE'}^S(f)
+ \alpha_3(f)\, \gamma_{EA'}^S(f)
}{
\sqrt{
\alpha_1^2(f) + \alpha_2^2(f) + \alpha_3^2(f)
}
},
\end{equation}
where the coefficients \( \alpha_i(f) \) are specific linear combinations of vector and tensor overlap reduction functions:
\begin{equation}\label{eq:alphas of before}
\begin{aligned}
\alpha_1(f) &= \gamma_{AE'}^V \gamma_{EA'}^T - \gamma_{EA'}^V \gamma_{AE'}^T, \\
\alpha_2(f) &= \gamma_{EA'}^V \gamma_{AA'}^T - \gamma_{AA'}^V \gamma_{EA'}^T, \\
\alpha_3(f) &= \gamma_{AA'}^V \gamma_{AE'}^T - \gamma_{AE'}^V \gamma_{AA'}^T.
\end{aligned}
\end{equation}

We also calculate the sensitivity curve, which is the equivalent energy density of the uncertainty,
\begin{equation}\label{eq:omega_nosie}
\Omega_{\mathrm{nosie}}^S(f)
= \frac{4\pi^2 f^3}{3(H_0/h)^2}
\biggl[
    \frac{\bigl|\Gamma^S(f)\bigr|^2}
         { N(f)\, N'(f)}
\biggr]^{-1/2}\,.
\end{equation}

To evaluate the sensitivity under different data reconstruction strategies, we consider four different data reconstructions between the LISA and Taiji detectors.  
\begin{itemize}
  \item \textbf{Reconstruction 1} corresponds to the combination used in this section, consisting of the channels \((A, A'), (A, E'), (E, A')\).
  \item \textbf{Reconstruction 2}  \((A, A'), (A, E'), (E, E')\).
  \item \textbf{Reconstruction 3}  \((A, A'), (E, A'), (E, E')\).
  \item \textbf{Reconstruction 4}  \((A, E'), (E, A'), (E, E')\).
\end{itemize}
Each reconstruction uses a specific linear combination of channels to isolate vector or scalar polarizations by doing ORF-weighted correlations. To isolate the vector mode, just swap $V$ and $S$ in every formula in Section 4. For any other reconstruction, swap the labellings of the channels in the correct order.

\begin{figure}
    \includegraphics[width=0.9\linewidth]{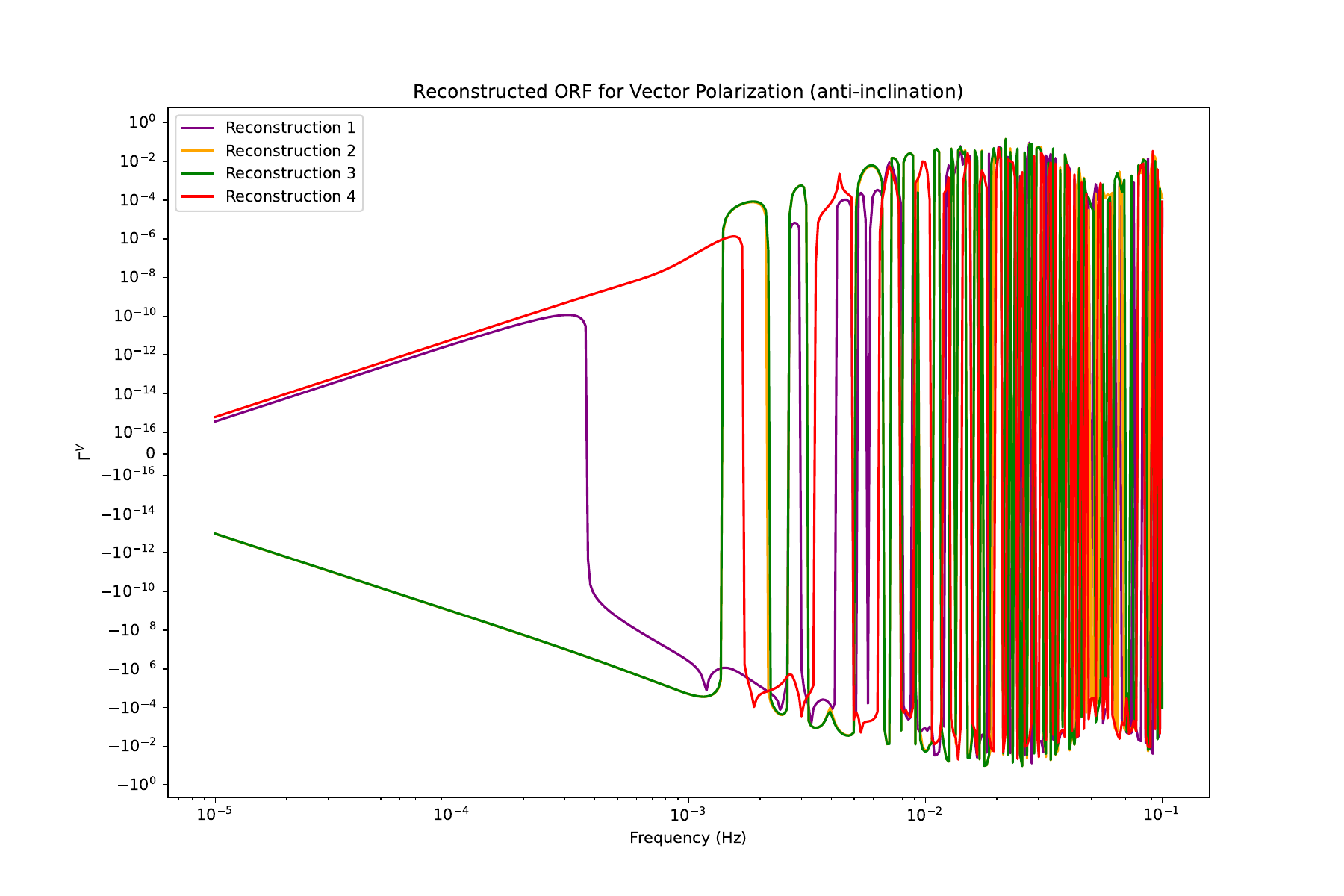}
    \includegraphics[width=0.9\linewidth]{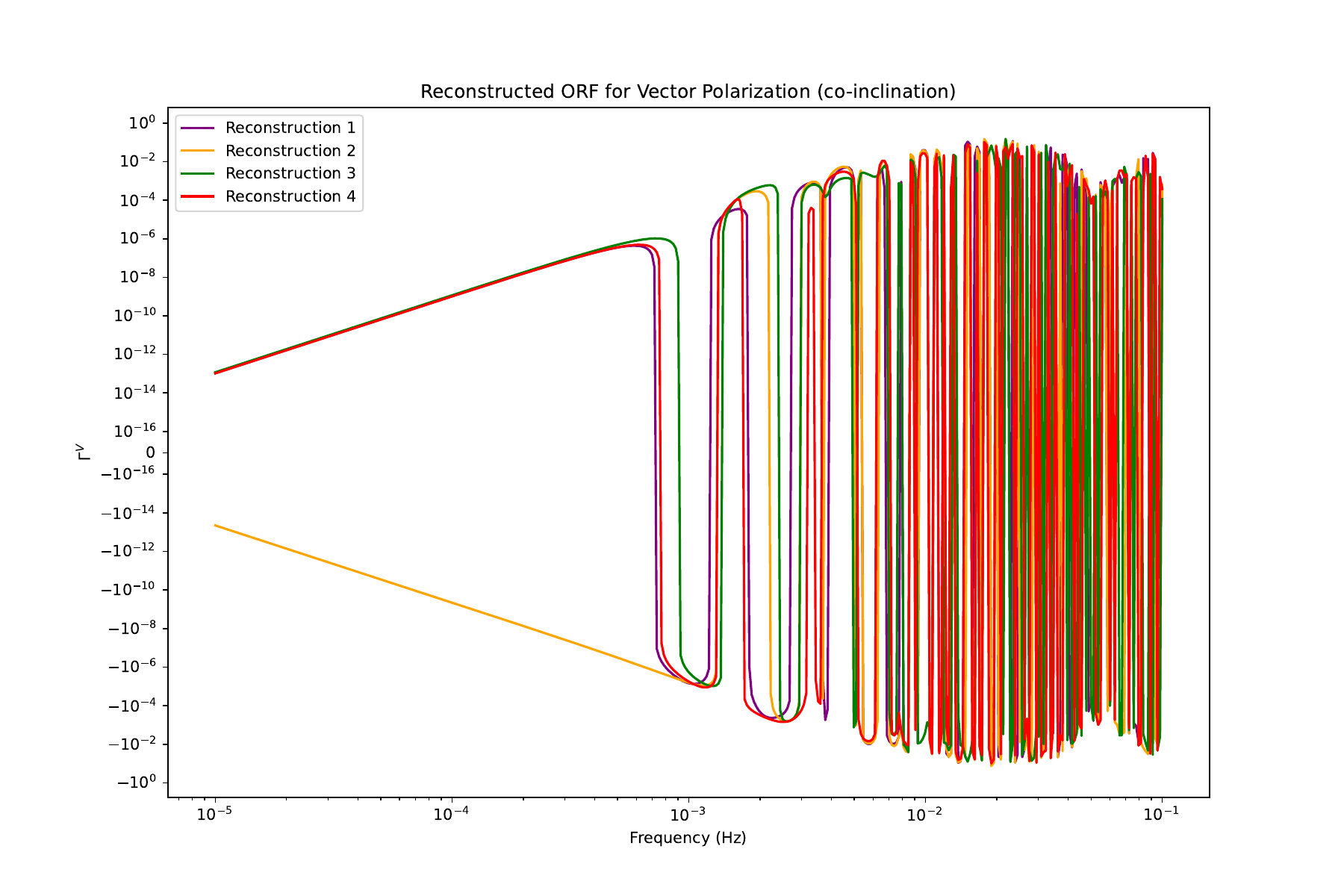}
    \includegraphics[width=0.9\linewidth]{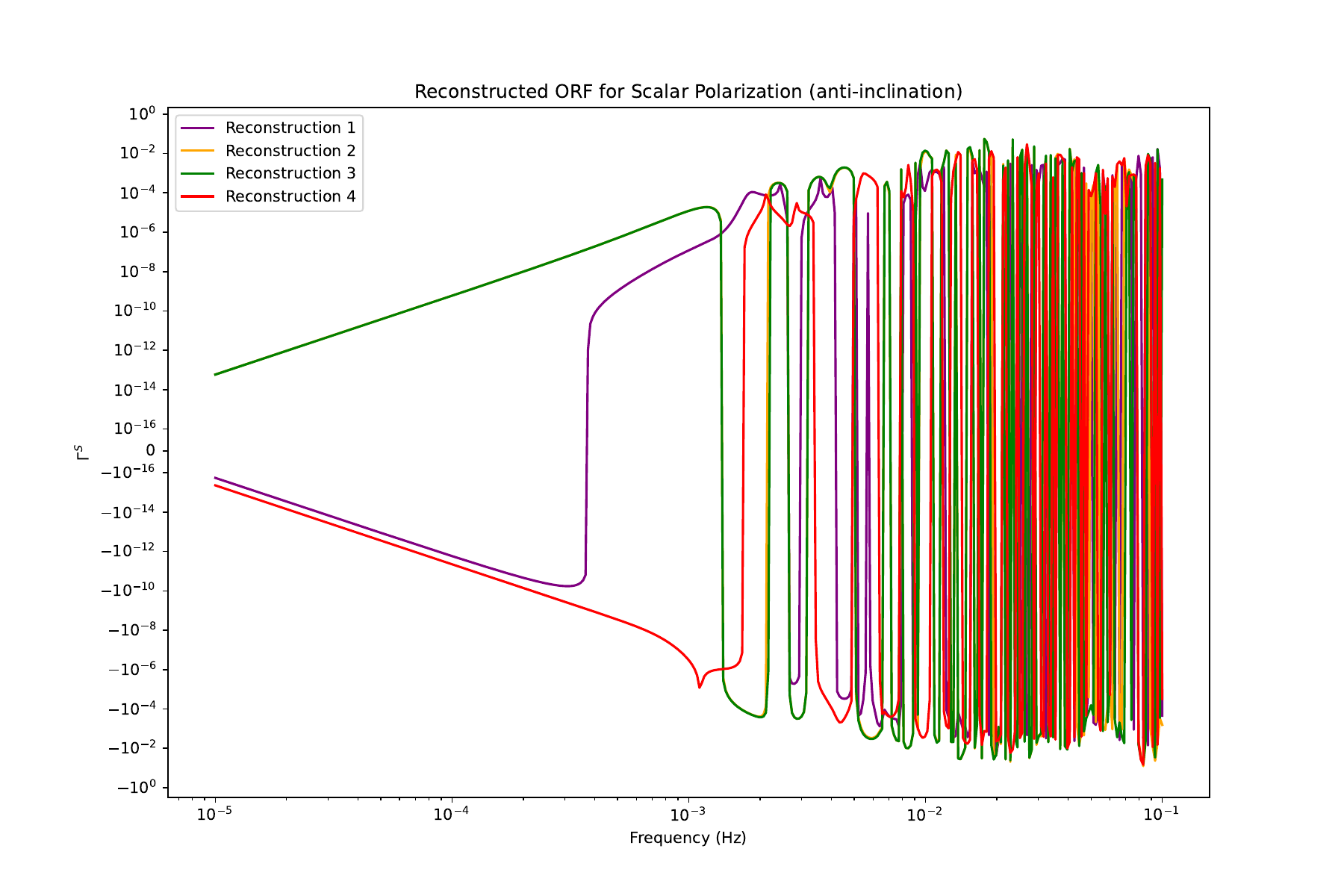}
    \includegraphics[width=0.9\linewidth]{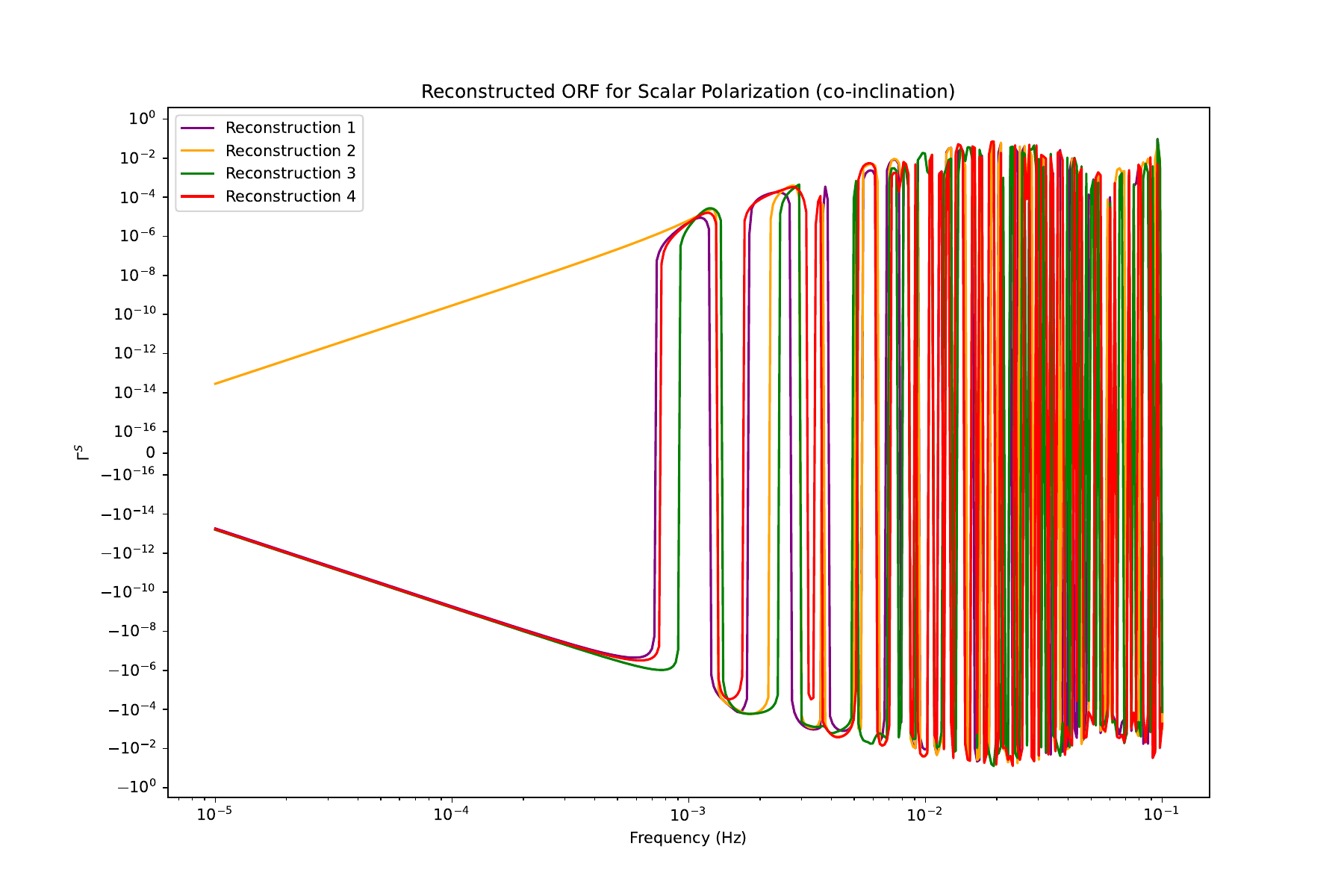}
    \caption{\textbf{The effective ORF $\Gamma$ of the reconstructed data.} The first picture shows the ORF for vector polarization of the anti-inclination LISA-Taiji network, the second one shows the ORF for vector polarization of the co-inclination network, the third one shows the ORF for scalar polarization of the anti-inclination network, and the last one shows the ORF for scalar polarization of the co-inclination network.}
    \label{fig:reconstructed ORF}
\end{figure}

\begin{figure}
    \includegraphics[width=0.9\linewidth]{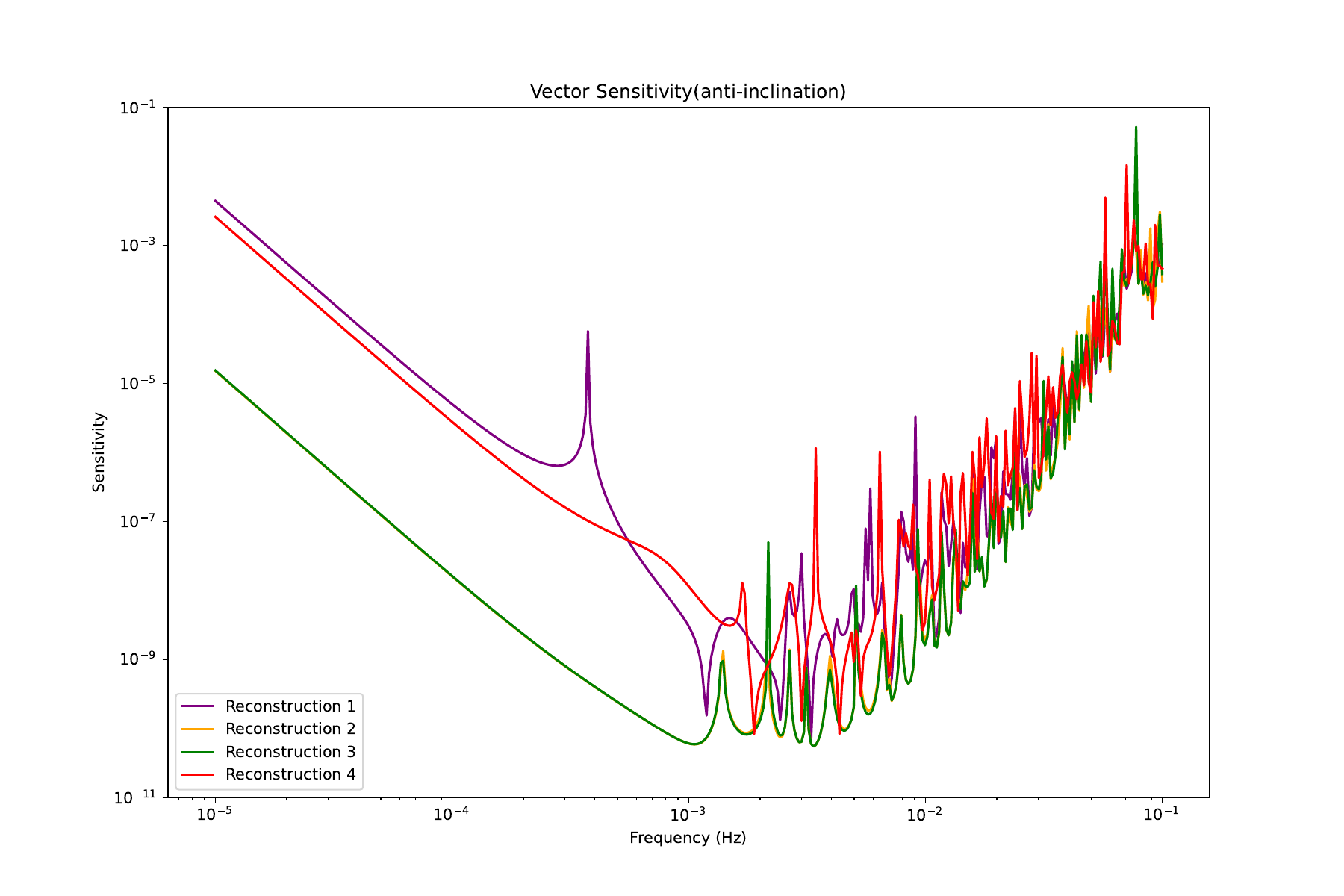}
    \includegraphics[width=0.9\linewidth]{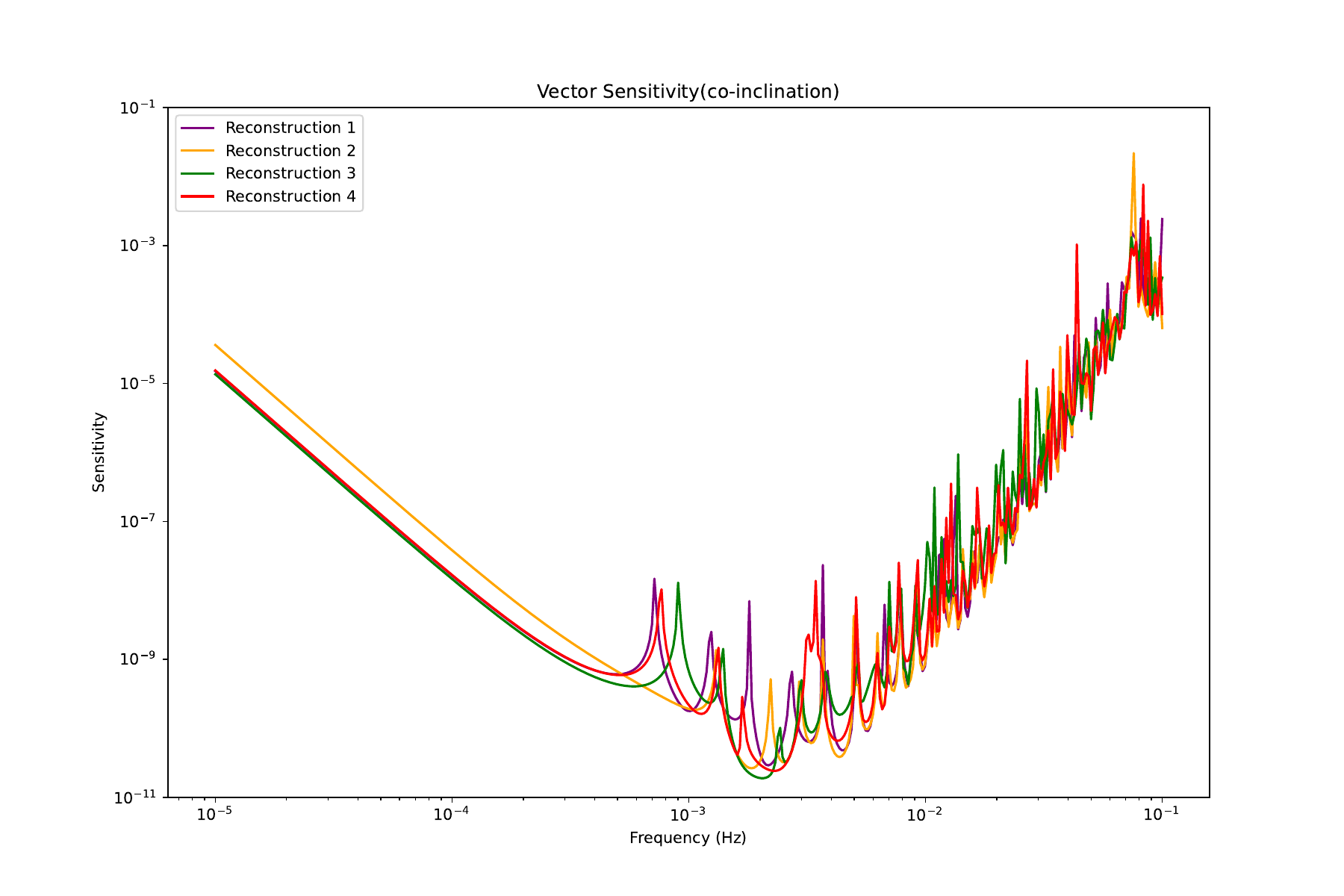}
    \includegraphics[width=0.9\linewidth]{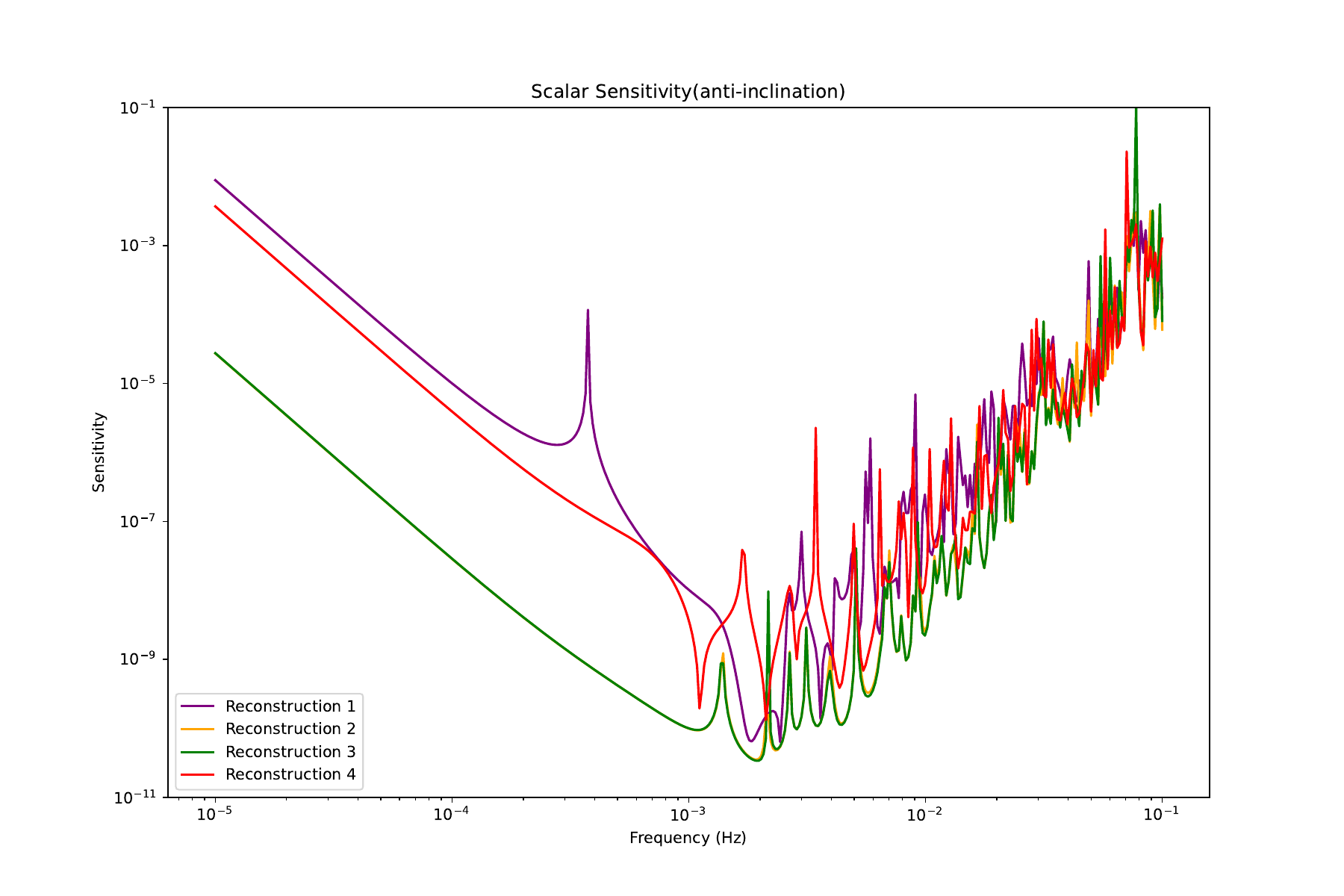}
    \includegraphics[width=0.9\linewidth]{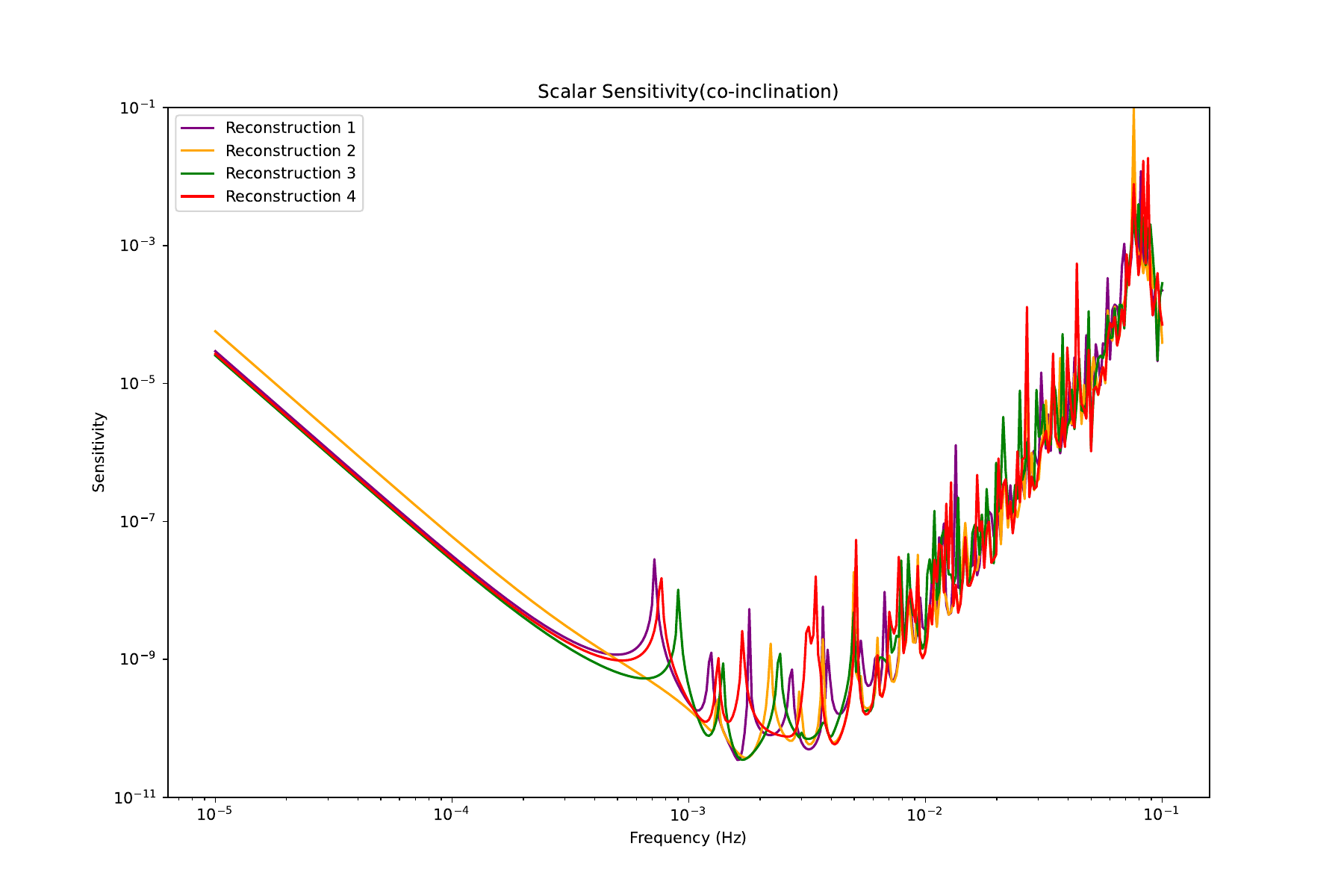}
    \caption{\textbf{The sensitivity curves of the reconstructed data.} The first picture shows the sensitivity curves for vector polarization of the anti-inclination LISA-Taiji network, the second one shows the sensitivity curves for vector polarization of the co-inclination network, the third one shows the sensitivity curves for scalar polarization of the anti-inclination network, and the last one shows the sensitivity curves for scalar polarization of the co-inclination network.}
    \label{fig:sensitivity curve}
\end{figure}

\emph{Results}. 
In this work, we draw the effective ORF $\Gamma$ of the reconstructed data in Fig.~\ref{fig:reconstructed ORF}, sensitivity curve in Fig.~\ref{fig:sensitivity curve} and calculate the SNRs of separated scalar and vector polarization components of SGWB using cross-correlation between LISA and Taiji. All the results for four reconstructions with two different Taiji configurations are shown as below.

We set the energy density of SGWB $\Omega_{GW}^p(f)=10^{-12}$ for both scalar and vector polarization components, which is a flat spectrum. The observation time is taken as $5$ years. For the vector mode, the SNRs for Reconstruction 1 to 4 with co-inclination LISA-Taiji network are 12.1, 15.6, 17.3, 15.3, with anti-inclination network are 2.1, 8.8, 9.0, 2.2.  For the scalar mode, the SNRs for Reconstruction 1 to 4 with co-inclination LISA-Taiji network are 8.8, 10.4, 9.8, 7.5, with anti-inclination network are 3.7, 9.5, 9.8, 1.1.

These results show that the SNRs are primarily determined by the sensitivity in the frequency range \(10^{-3} \sim 10^{-2}\,\mathrm{Hz}\). The reconstruction group that exhibits the best sensitivity performance within this frequency band achieves the highest total SNR. In Fig.~\ref{fig:sensitivity curve}, the co-inclination network shows better sensitivity than the anti-inclination network. In conclusion, for both vector and scalar polarization separation of isotropic SGWB, the co-inclination network outperforms the anti-inclination one. Fig.~\ref{fig:reconstructed ORF} shows the same conclusion, as co-inclination network provides larger effective ORFs than anti-inclination network.

\emph{Discussions}. 
Our results demonstrate that isolating individual polarization modes of an isotropic SGWB via ORF‐weighted data reconstruction in the LISA–Taiji network is fully viable in practice. With observing time of the LISA-Taiji network reaching 10 years, both Taiji orbital configurations can reach the detection limit of $\Omega_{GW}^p(f)\sim10^{-12}$ with SNRs exceeding $10$. Since isolating particular polarization mode is possible, we can be sure about which polarization mode exists or not. With this existence check on vector mode and scalar mode, we can distinguish gravity theories much better than before.

When calculating SNRs, we assume that $\Omega^p_{GW}(f)$ is a flat spectrum. This is a convenient choice for SNR calculation, and can approximately show the overall detectability of the reconstructed data. When doing real data analysis, our procedure can handle any spectrum form for $\Omega^p_{GW}(f)$. If a different $\Omega^p_{GW}(f)$ is taken, as long as its amplitude is similar to our choice in the most sensitive frequency band of LISA-Taiji network, the detectability to that spectrum form is guaranteed as well.

Screening mechanisms in modified gravity theories, such as chameleon- or
Vainshtein-type screening, mainly suppress extra degrees of freedom in
high-density or strongly self-gravitating environments and thus affect the
generation of non-tensorial gravitational radiation at the source
\cite{KhouryWeltman2004,HinterbichlerKhoury2010,Zhang:2016hwu,Shao2017}.
Once generated, gravitational waves propagate in the wave zone where the
perturbations are well described by linearized field equations, and screening
effects are expected to be negligible.
Since the response of space-based interferometers depends only on the linear
coupling between metric perturbations and freely falling test masses, screening
does not modify the overlap reduction functions or the polarization separation
method developed in this work.

However, we emphasize that ORFs are not static. They evolve continuously as the spacecraft move along their orbits and depend on the specific labeling of the spacecraft indices.
Ref.~\cite{wang2025testgravitationalwavepolarizationsspacebased} shows that phase difference between LISA and Taiji orbitals doesn't affect the GW detectability, but it does influence the ORF, which represents the SGWB detectability. The ORF curves presented in this article serve as representative examples; while their overall magnitudes remain stable, the detailed numerical values will shift over time. In real‐data analyses, one must therefore recompute the ORFs every day or even every hour to match the exact constellation geometry at each observation epoch. This requirement does not impose a significant computational burden, since ORF evaluation is already a standard part of any SGWB search pipeline of LISA-Taiji network. Besides, with the expected value and variance our reconstruction method provides, it is very simple to construct a likelihood function to match a SGWB search pipeline.

In this article, we assumed that $b, L$ polarizations have the same statistics like the corresponding power spectral density, but some theories may disagree. Thus the need of further polarization separation rises. With four usable cross-correlation data channels in the LISA-Taiji network, it is also possible to separate the $b,L$ polarizations by extending our method.

Some previous studies~\cite{OmiyaSeto2020} have argued that the co-inclination Taiji configuration cannot achieve full polarization separation. This result rose from the choice of coordinate and calculation of ORFs. Omiya \& Seto rotated the strain data of both space-based detectors to a special angle, and that special angle yields zero ORFs for two cross-correlation channels. In this work, our ORF calculation method guarantees that the co‐inclination network still yields valid ORFs with AE' and EA' cross-correlation channels.

Finally, earlier analyses~\cite{WangNiHan2021,Wang2021} asserting that the anti‐inclination configuration outperforms the co‐inclination one when detecting non-GR polarization modes were focused on the polarization recovery for individual GW events, which highly depends on the angular resolutions. In contrast, the sensitivity of separated polarization signal of an isotropic SGWB only relies on the overlap-reduction functions. Thus, there is no true contradiction: both approaches agree that the anti‐inclination network offers a stronger anisotropic detectability, while the co‐inclination network remains a perfectly acceptable alternative for statistical background measurements.

	%%%%%%%%%%%%%%%%%%%%%%%%%%%%%%%%%%%%%%%%%%%%%%%%%%%%%%%%%%%%%%%%%
	\emph{Acknowledgments}
	We thank Gang Wang and Minghui Du for his valuable discussions and advice. This work is supported in part by the National Natural Science Foundation of China under Grant No.12235019 and No.12475067.

	\bibliographystyle{apsrev}
	\bibliography{ms}
\end{document}